\documentclass[a4paper,11pt]{article}
\pdfoutput=1
\usepackage{jheppub}
\usepackage[T1]{fontenc}
\usepackage[utf8]{inputenc}
\usepackage{microtype}
\usepackage{lmodern}
\usepackage{booktabs}
\usepackage{mdframed}
\usepackage{graphicx}
\usepackage{verbatim}
\usepackage{hyperref}
 \usepackage{xcolor}
 \definecolor{dark-red}{rgb}{0.4,0.15,0.15}
 \definecolor{dark-blue}{rgb}{0.15,0.15,0.4}
 \definecolor{medium-blue}{rgb}{0,0,0.5}
 \hypersetup{colorlinks, linkcolor={dark-red}, citecolor={dark-blue}, urlcolor={medium-blue}}
\usepackage{tikz}
\usetikzlibrary{arrows.meta, decorations.pathmorphing, positioning}

\usepackage{amsmath,amsfonts,amssymb}
\usepackage{eucal}
\usepackage{mleftright}
\usepackage{tensor}
\usepackage{braket}
\usepackage{bm}
\usepackage{mathtools}
\usepackage{mathrsfs}
\newcommand{\mL}{\mathcal{L}}

\newcommand{\mO}{\mathcal{O}}

\newcommand{\mJ}{\mathcal{J}}
\newcommand{\mM}{\mathcal{M}}
\newcommand{\mU}{\mathcal{U}}
\newcommand{\mT}{\mathcal{T}}
\newcommand{\mW}{\mathcal{W}}

\newcommand{\circColon}{\mathrel{\substack{\circ \\[-0.3ex] \circ}}}

\newcommand{\vdotslower}{\,\raisebox{-0.4ex}{$\vdots$}\,}
\newcommand{\triplenormord}[1]{\mathopen{\vdotslower}#1\mathclose{\vdotslower}}

\title{\boldmath Free-field construction of Carrollian $W_N$-algebras}

\author[a]{Stefan Fredenhagen}
\author[b]{and Lucas H{\"o}rl}

\affiliation[a]{University of Vienna, Faculty of Physics, Mathematical Physics, Boltzmanngasse 5, 1090, Vienna, Austria}
\affiliation[b]{Institute for Theoretical Physics, TU Wien, Wiedner Hauptstrasse 8-10, 1040 Vienna, Austria}

\abstract{We study Carrollian contractions of $W_N$-algebras from a free-field perspective. Using a contraction of the Miura transformation, we obtain explicit free-field realizations of the resulting Carrollian $W_N$-algebras. At the classical level, they are isomorphic to the Galilean $W_N$-algebras. In the quantum case, we distinguish between two Carrollian constructions: a flipped Carrollian contraction, where the time direction is reversed in one sector, and a symmetric contraction. The flipped construction yields a quantum algebra isomorphic to the Galilean one, whereas the symmetric construction produces a distinct quantum Carrollian $W_N$-algebra whose basic structure constants are identical to those of the classical Carrollian $W_N$-algebra. These algebras provide a natural framework for studying extended symmetries in Carrollian conformal field theories, motivated by recent developments in flat space holography. Our construction provides tools for developing the representation theory of Carrollian (and Galilean) $W_N$-algebras using free-field techniques.
}

\begin{document} 
\maketitle

\section{Introduction}
\label{sec:intro}

Carrollian symmetries \cite{LevyLeblond:1965} have recently emerged as central in the exploration of non-Lorentzian limits of relativistic theories \cite{Barnich:2012rz,Duval:2014uoa,Henneaux:2021yzg,deBoer:2021jej, Rivera-Betancour:2022lkc, Baiguera:2022lsw, Miskovic:2023zfz}, with wide-ranging applications in areas such as flat space holography \cite{Bagchi:2010zz,Bagchi:2012cy,Barnich:2012aw,Duval:2014uva,Bagchi:2016bcd,Bagchi:2022emh,Donnay:2022aba,Nguyen:2023vfz,Nguyen:2023miw,Bagchi:2023cen}, tensionless and null strings \cite{Bagchi:2013bga,Bagchi:2015nca}, black hole physics \cite{Donnay:2019jiz,Penna:2018gfx, Ecker:2023uwm}, cosmology \cite{deBoer:2021jej, deBoer:2023fnj}, hydrodynamics \cite{deBoer:2017ing,Ciambelli:2018xat,Petkou:2022bmz,Bagchi:2023ysc}, the fluid/gravity correspondence \cite{Ciambelli:2018wre, Campoleoni:2018ltl, Campoleoni:2022wmf}, and condensed matter systems \cite{Bidussi:2021nmp,Bagchi:2022eui}. The associated Carroll algebra arises as an ultra-relativistic Inönü--Wigner contraction of the Poincaré group, corresponding to the formal limit in which the speed of light is sent to zero \cite{LevyLeblond:1965,SenGupta:1966qer}. At the level of conformal symmetries, this contraction yields the $d$-dimensional Carrollian conformal algebra, which coincides with the BMS$_{d+1}$ algebra, the asymptotic symmetry algebra of
$(d+1)$-dimensional asymptotically flat spacetimes \cite{Bondi:1962px,Sachs:1962zza,Barnich:2006av}. 

The recognition that the BMS$_3$ algebra is isomorphic to a contraction of two copies of the Virasoro algebra~\cite{Bagchi:2010zz} has provided a conceptual foundation for \emph{flat space holography}, proposing that gravity in asymptotically flat spacetime admits a dual description in terms of a codimension-one Carrollian conformal field theory (CCFT) living at null infinity \cite{Bagchi:2023cen}.\footnote{Another approach is celestial holography \cite{Strominger:2017zoo}, which provides a complementary perspective on the flat-space scattering sector \cite{Donnay:2022aba,Bagchi:2022emh}.} Besides studying supersymmetric extensions (see for example \cite{Bagchi:2009ke, Bagchi:2022owq, Lodato:2016alv, Barnich:2015sca}), a natural generalization of such a holographic duality is to include higher-spin fields in the bulk. In AdS$_3$, higher-spin gravity is known to exhibit asymptotic $W$-symmetries~\cite{Campoleoni:2010zq,Henneaux:2010xg}, suggesting that its holographic duals are 2D CFTs with extended conformal symmetry generated by $W$-algebras. The flat-space analogue of this structure was investigated in~\cite{Afshar:2013vka,Gonzalez:2013oaa}, where a spin-3 gravity theory in flat space was shown to possess a Carrollian version of the $W_3$-algebra as its asymptotic symmetry. More generally, one expects that a flat space higher-spin gravity theory (based on a corresponding flat higher-spin algebra~\cite{Ammon:2017vwt,Campoleoni:2021blr}) has an asymptotic symmetry given by the corresponding \emph{Carrollian $W$-algebra}. It arises from a Carrollian contraction of the relativistic $W$-algebras\footnote{Similar to the contraction of Galilean $W$-algebras \cite{Rasmussen:2017}.} and extends the Carrollian conformal algebra (or BMS$_3$ algebra) by higher-spin currents, naturally generalizing the Carrollian symmetry paradigm.

In two dimensions, the Carrollian conformal algebra is isomorphic to the Galilean conformal algebra, which is intuitively connected to a duality exchanging time and the one-dimensional space. Similarly, Carrollian and Galilean contractions of $W_N$-algebras yield equivalent higher-spin extensions at the classical level \cite{Campoleoni:2016vsh}. However, as noted in~\cite{Campoleoni:2016vsh}, at the quantum level the situation is more delicate because the nonlinearities in the algebras for $N>2$ require a notion of normal ordering which itself is connected to the identification of a time direction, and it is thus plausible that the naïve exchange of time and space fails in the nonlinear case. To make this explicit, we study on the one hand a flipped Carrollian contraction, in which the time direction in one sector is formally reversed, leading to an algebra isomorphic to that obtained from a Galilean contraction. On the other hand, we consider a symmetric Carrollian contraction, which keeps the time direction untouched and should thus be regarded as the proper Carrollian contraction. 

In relativistic two-dimensional CFTs, $W_N$-algebras have a rich history as symmetry algebras of integrable models and coset theories \cite{Bouwknegt:1992wg}. The $W_3$-algebra was first considered as an extension of the Virasoro algebra by a holomorphic primary current of spin 3 \cite{Zamolodchikov:1985wn}. It was then recognized~\cite{Fateev:1987zh} that the higher-spin currents in $W_N$-algebras can be systematically realized as composite fields of free bosons via the \emph{quantum Miura transformation}. The free-field realization provides an efficient method for encoding the nonlinear structure of $W$-algebras in terms of simpler underlying fields, and has been widely used in both the classical~\cite{Dickey:1997wia}, and quantum cases~\cite{Fateev:1987zh,Bouwknegt:1992wg,Zhu:1993mc}. These realizations are particularly well suited for constructing Verma modules, analyzing null vectors, and computing correlation functions, especially in rational or coset-based models~\cite{Bouwknegt:1992wg}. This approach should remain fruitful after a Carrollian (or a Galilean) contraction.

In this work, we take a step in this direction by constructing Carrollian $W_N$-algebras via free fields. Starting from two copies of the relativistic $W_N$-algebra, we apply an ultra-relativistic contraction at the level of the Miura operator, obtaining explicit expressions for the higher-spin currents of Carrollian $W_N$-algebras. Specifically, the Carrollian $W_N$-algebra is generated by two sets of fields $\mU^+_k$ and $\mU^-_k$, where $k=2,\dots,N$. They can be implicitly expressed as products of free fields $\mJ^\pm_i$ ($i=1,\dots,N)$ satisfying $\sum_i \mJ^\pm_i=0$ (where all $\mJ^+_i$ commute, as well as all $\mJ^-_i$, and $\mJ^+_i$ and $\mJ^-_j$ have a non-trivial commutator given in~\eqref{PoissonbracketsJi}) via the differential operators 
\begin{align}
    (\alpha^+ \partial - \mJ^+_1) \cdots (\alpha^+ \partial -\mJ^+_N) &= (\alpha^+ \partial)^N - \sum_{k=2}^N \mU^+_k(\alpha^+ \partial)^{N-k}\,,\\
     \sum_{j=1}^N (\alpha^+\partial -\mJ_1^+) \cdots \mJ_j^-  \cdots (\alpha^+\partial - \mJ^+_N) &= \sum_{k=2}^N \big(\mU^-_k -\alpha^- \partial_{\alpha^+}\mU^+_k\big) (\alpha^+\partial )^{N-k}, 
     \label{CarrollMiura2}
\end{align}
depending on two parameters $\alpha^\pm$. Here, the summands on the left-hand side of the second equation are understood as a product of factors $(\alpha^+\partial - \mJ^+_i)$ where the $j^\mathrm{th}$ factor is replaced by $\mJ^-_j$. The first equation determines $\mU^+_k$ as functions of $\alpha^+$, and in the second, their derivatives with respect to $\alpha^+$ are needed to determine $\mU^-_k$. 

In the quantum case, the expression on the left-hand side of~\eqref{CarrollMiura2} needs an ordering prescription. By choosing the usual normal ordering for the currents $\mJ^\pm_i$, one obtains the Galilean version of the quantum $W_N$-algebra with central charges
\begin{align}
    \tilde{c}_L &= (N-1)\big(2-4 N(N+1)\alpha^+ \alpha^-\big)\,, & \tilde{c}_M &= -2 (N-1)N(N+1)(\alpha^+)^2\,.
\end{align}
For a proper Carrollian version, one needs to adopt an ordering for the left-hand side of~\eqref{CarrollMiura2} that averages the standard normal ordering and the anti-normal ordering. The corresponding central charges are
\begin{align}
    c_L &= -4 (N-1) N(N+1)\alpha^+ \alpha^-\,, & c_M &= -2 (N-1)N(N+1)(\alpha^+)^2\,,
\end{align}
which coincide with the classical expressions. A schematic overview over the resulting structures can be found in figure~\ref{fig1}.

\begin{figure}
\begin{center}
\begin{tikzpicture}[>=Latex, font=\footnotesize]

\node[draw, rounded corners, align=center] (rel) at (1,2.6) {\textbf{classical}\\ \textbf{relativistic $\boldsymbol{W_N\otimes W_N}$-algebra}};

\node[draw, rounded corners, align=center] (car) at (1,0) {\textbf{classical}\\ \textbf{Carrollian/Galilean}\\ \textbf{$\boldsymbol{W_N}$-algebra}};

\draw[->, bend right=5] (rel) to node[left=-1pt, align=center]
      {\scriptsize Carrollian\\[-4pt] \scriptsize contraction\ \ } (car);

  \draw[->, bend left=5] (rel) to node[right=-8pt, align=center]
      {\scriptsize Galilean\\[-4pt] \scriptsize \ \ \ contraction} (car);

  \node[draw, rounded corners, align=center] (relQ) at (8,2.6) {\textbf{quantum}\\ \textbf{relativistic $\boldsymbol{W_N\otimes W_N}$-algebra}};

\node[draw, rounded corners, align=center] (carQ) at (5.7,-0.2) {%
  \textbf{quantum}\\
  \textbf{Carrollian}\\
  \textbf{$\boldsymbol{W_N}$-algebra}\\
  [2pt]\textit{averaged ordering}\\[-1pt]
  {\fontsize{6.5pt}{10.8pt}\selectfont (classical structure constants)}
};
  
 \node[draw, rounded corners, align=center] (galQ) at (10.3,-0.2)  {\textbf{quantum}\\ \textbf{Galilean}\\ \textbf{$\boldsymbol{W_N}$-algebra}\\
  [2pt] \textit{normal ordering}\\ [-2pt]
  {\fontsize{6.5pt}{10.8pt}\selectfont (quantum structure constants)}
  };

  \draw[->, bend right=24] (relQ) to node[left=-12pt, align=center]
      {\scriptsize Carrollian\\[-4pt] \scriptsize contraction\ \ \ \ \ \ \ \ } (carQ);

  \draw[->, bend left=24] (relQ) to node[right=-20pt, align=center]
      {\scriptsize Galilean\\[-4pt] \ \ \ \ \ \ \ \ \scriptsize contraction} (galQ);

  \draw[->, dashed, bend left=20] (relQ) to
      node[left=-5pt, align=center]
      {\scriptsize flipped\ \ \ \ \ \ \\[-4pt] \scriptsize Carrollian} (galQ);

  \draw[->, dashed, bend right=20] (relQ) to
      node[right=-8pt, align=center, text=gray]
       {\ \ \ \ \  \scriptsize flipped\\[-4pt] \scriptsize Galilean} (carQ);
       
\end{tikzpicture}
\end{center}
\caption{\label{fig1}Whereas Carrollian and Galilean contraction of the $W$-algebras lead to isomorphic algebras in the classical case (related by exchange of space and time), one finds two different structures for the quantum algebras. Quantum analogues of the classical isomorphism are the isomorphism between the algebra obtained by the Galilean contraction and by a flipped Carrollian contraction (where normal ordering is flipped in one sector), and, analogously the isomorphism between the Carrollian contraction and a flipped Galilean one (not discussed in this work). The two quantum algebras differ in the ordering prescription and in the structure constants, which only in the case of the quantum Galilean algebra receive quantum corrections.}
\end{figure}

By expressing the $W_N$-currents as combinations of free-field currents, our construction opens the door to building and investigating Carrollian (or Galilean) representations through free-field techniques. This can serve as a foundational tool for investigating physical models with extended Carrollian symmetry, including putative duals in flat space holography.
\smallskip 

This paper is organized as follows: In Section 2, we review the Carrollian conformal algebra and introduce two free-field realizations, one that is based on the usual normal ordering (a Galilean free-field realization) and one that is based on averaged ordering (a Carrollian free-field realization). In Section 3, we construct the classical Carrollian $W_N$-algebras via a Carrollian Miura transformation and provide explicit expressions in the example $N=2$. Section 4 extends this to the quantum case, highlighting the subtleties arising from the ordering prescription. We discuss a flipped and a symmetric construction of commutator algebras whose classical version is a Carrollian $W_N$-algebra. The algebra arising from the flipped construction is isomorphic to the quantum Galilean $W_N$-algebra, whereas the symmetric one is a proper Carrollian $W_N$-algebra. We conclude in Section 5 with a brief discussion of possible applications and future directions. In addition to a technical remark on reversed normal ordering, the appendix contains a discussion of the Galilean contraction of the Miura transformation and its relation to the Carrollian one, as well as an illustration of the free-field construction of the Galilean $W_3$-algebra in terms of its basic operator product expansions.

\section{The Carrollian conformal algebra and its free-field realization}
\label{sec:CCA}

This section reviews the two-dimensional Carrollian conformal algebra that arises from a contraction of two copies of the Virasoro algebra. We present two free-field constructions of this algebra, a Galilean and a Carrollian one, for generic values of the central charges.

\subsection{Carrollian contraction and Carrollian conformal algebra}

Starting from the Poincaré algebra, one obtains the Carroll algebra as an Inönü--Wigner contraction in which the speed of light is scaled to zero \cite{LevyLeblond:1965, Bacry:1968zf}. More concretely, under the rescaling $c \to c\epsilon^2$, only the boost generators $B_i = - c^2 t \partial_i - x_i \partial_t$ are affected. In the vanishing speed of light limit $\epsilon\to 0$, the Hamiltonian $H$ becomes a central element of the Carroll algebra. Carroll boosts $b_i = -x_i \partial_t$ generally commute, a significant distinction compared to the Poincaré algebra. The generators of the Carroll algebra are time translations, spatial translations, Carroll boosts and spatial rotations:
\begin{align}
    H&=\partial_t\,, & P_i &=\partial_i\,,& b_i &= -x_i \partial_t\,,& J_{ij} &= x_i \partial_j - x_j \partial_i\,.
\end{align}
Only the commutators that involve the boosts are modified relative to the Poincaré algebra:
\begin{align}
    [b_i, b_j] &= 0\,, & [b_i, H] &= 0\,,&  [b_i,P_j] &= \delta_{ij} H\,, &  [b_i, J_{kl}] &= \delta_{ik} b_l - \delta_{il} b_k\,.
\end{align}
The other commutators
\begin{align}
    [P_i,P_j]&=0\,,& [J_{ij}, J_{kl}] &= \delta_{ik} J_{jl}+ \delta_{jl} J_{ik} -\delta_{il} J_{jk} -\delta_{jk} J_{il}\,,&  [J_{ij}, P_k] &= \delta_{jk} P_i-\delta_{ik} P_j\,,
\end{align}
are unaffected. Carrollian theories reverse the roles of space and time compared to Galilean spacetimes. Consequently, finite Carroll boosts leave space invariant, but change time
\begin{align}
     \tilde{x}^i &= x^i\, & \tilde{t}&=t-\lambda_i x^i\,,
\end{align}
the exact opposite of Galilean transformations. See the review \cite{Bagchi:2025vri} for details.

A similar contraction can be performed for conformal symmetries. In the particular case of two dimensions, conformal symmetry gives rise to two copies of the Virasoro algebra, with the Lie bracket
\begin{align} \label{Virasoro-algebra}
    [L_m, L_n] = (m-n)L_{m+n} + \frac{c}{12} m(m^2-1)\delta_{m+n,0},
\end{align}
and analogously for the generators $\bar{L}_m$ of the second copy. A Carrollian contraction can then be defined by considering the combinations
\begin{align}
    \mathcal{L}_m &= L_m - \bar{L}_{-m}\,,& \mathcal{M}_m &= \epsilon^2(L_m + \bar{L}_{-m})\,,
    \label{Carroll-Virasoro-trafo}
\end{align}
where $\epsilon$ denotes the Carrollian contraction parameter from above.
Evaluating the Lie brackets in the limit $\epsilon\to 0$, one obtains the Carrollian conformal algebra with commutation relations
\begin{subequations}\label{CCA}
\begin{align}
    [\mathcal{M}_m,\mathcal{M}_n]&=0\,, 
    \label{N=2-commutators-p1}\\
    [\mathcal{L}_m,\mathcal{L}_n] &= (m-n)\mathcal{L}_{m+n} + \frac{c_L}{12}m(m^2-1)\delta_{m,-n}\,,\\
    [\mathcal{L}_m,\mathcal{M}_n] &= (m-n)\mathcal{M}_{m+n} + \frac{c_M}{12}m(m^2-1)\delta_{m,-n} \,.
    \label{N=2-commutators-p3}
\end{align}
\end{subequations}
Note that these are also the commutator relations of the Galilean conformal algebra or BMS$_3$ algebra. In two dimensions, this is intuitive: the Carrollian limit $\epsilon\to0$ corresponds to the light cone collapsing along the time axis, whereas the Galilean limit $\epsilon\to\infty$ corresponds to collapse along the spatial axis. Exchanging the role of space and time gives rise to an isomorphism of the corresponding symmetry algebras. 
\subsection{Free-field construction}\label{sec:freefieldconstruction}

As is well known, the Virasoro algebra can be realized in terms of the modes of a free field. This can be understood by considering the energy-momentum tensor of a free field in the presence of a background charge. In the holomorphic sector, we then have 
\begin{equation}\label{freefieldconofT}
    T=\, :\!JJ\!:+\alpha \,\partial J\,,
\end{equation} 
where $T(z)=\sum_n L_n z^{-n-2}$ is the holomorphic component of the energy-momentum tensor, and $J(z)=\sum_n J_n z^{-n-1}$ is the conserved holomorphic spin-1 current of a free field.\footnote{Here we use a normalization where the singular operator product expansion of $J(z)$ reads $J(z)J(w)\sim \frac{1}{2}(z-w)^{-2}$, in order to be consistent with our discussion later.} The colons denote the standard normal ordering on the plane, which for a free spin-1 field simply means placing generators $J_m$ with positive mode numbers to the right. In terms of modes one obtains
\begin{equation}
    L_m = \sum_n :\!J_{m-n} J_n\!: - (m+1)\alpha J_m\,,
\end{equation}
which provides a construction of the Virasoro algebra with central charge $c=1-6\alpha^2$ in terms of the oscillators of a free field, satisfying the commutation relations
\begin{equation}\label{freefieldCR}
    [J_m,J_n] = \frac{m}{2}\,\delta_{m,-n}\,.
\end{equation}
Similarly, one can realize the Carrollian conformal algebra in terms of free fields as we present below. We start from generators $\mJ^\pm_m$ with commutation relations
\begin{align}
    [\mJ^+_m,\mJ^-_n] &= \frac{m}{4}\,\delta_{m,-n}\,, & [\mJ^+_m,\mJ^+_n]&=0\,,& [\mJ^-_m,\mJ^-_n]&=0\,.
\end{align}
We then define
\begin{subequations}\label{FreefieldconofCCA}
\begin{align}
    \tilde{\mL}_m &=  4\sum_n :\!\mJ^+_{m-n}\mJ^-_n \!:- 2(m+1)\big(\alpha^+ \mJ^-_m + \alpha^-\mJ^+_m\big)\,,\\
    \tilde{\mM}_m &= 2\sum_n \mJ^+_{m-n}\mJ^+_n  - 2(m+1)\alpha^+\mJ_m^+\,,
\end{align}
\end{subequations}
where the normal ordering $:\ :$ symbol again indicates that generators $\mJ^\pm_m$ with positive mode numbers are shifted to the right. Note that normal ordering is only needed in the definition of $L_0$. One can then explicitly check that the commutation relations~\eqref{CCA} of the Carrollian conformal algebra are satisfied with central charges\footnote{For $\alpha^+ =\alpha^- =0$, this construction corresponds to the quantization of one free scalar with respect to the 'flipped vacuum' \cite{Bagchi:2020fpr}.}
\begin{align}\label{Carroll-central-charges-N=2}
    \tilde{c}_L &= 2-24\alpha^+\alpha^-\,,& \tilde{c}_M&= -12 (\alpha^+)^2\,.
\end{align}
There is a different free-field construction of the Carrollian conformal algebra that is not based on normal ordering, but on an averaged ordering,
defined as the average of normal and anti-normal ordering,
\begin{equation}\label{symmordering}
    \triplenormord{\mJ^+_m \mJ^-_{-m}} = \frac{1}{2} \big( \mJ^+_m \mJ^-_{-m} + \mJ^-_{-m} \mJ^+_m \big)\,.
\end{equation}
Similarly to~\eqref{FreefieldconofCCA}, we define the generators $\mL_m$, $\mM_n$ in terms of free fields, 
\begin{subequations}\label{2ndFreefieldconofCCA}
\begin{align}
    \mL_m &=  4\sum_n \triplenormord{\mJ^+_{m-n}\mJ^-_n} - 2(m+1)\big(\alpha^+ \mJ^-_m + \alpha^-\mJ^+_m\big)\,,\\
    \mM_m &= 2\sum_n \mJ^+_{m-n}\mJ^+_n  - 2(m+1)\alpha^+\mJ_m^+\,.
\end{align}
\end{subequations}
One can check in a straightforward calculation that they satisfy the Carrollian conformal algebra~\eqref{CCA} with central charges
\begin{align}\label{Carroll-central-charges-N=2-classical}
    c_L &= -24\alpha^+\alpha^-\,,& c_M&= -12 (\alpha^+)^2\,,
\end{align}
which coincide with the classical central charges.\footnote{The constant shift in $\tilde{c}_L$ (see~\eqref{Carroll-central-charges-N=2}) stems from normal ordering.}

We refer to the second construction as the \emph{symmetric} free-field construction, whereas we call the first construction the \emph{flipped} construction. Both constructions allow for arbitrary central charges as long as $c_M\not=0$. In $\mathrm{AdS}_3/\mathrm{CFT}_2$, one finds for Einstein gravity the Brown-Henneaux central charges $c=\bar c = \frac{3l}{2G_\mathrm{N}}$ \cite{Brown:1986nw} of the asymptotic symmetry algebra, where $G_\mathrm{N}$ is Newton’s constant and $l$ is the AdS radius. The flat limit then corresponds to sending $l\to\infty$. A transformation as in \eqref{Carroll-Virasoro-trafo}, where $1/l$ essentially plays the role of the Carrollian contraction parameter $\epsilon$, can be used to obtain the $3\mathrm{d}$ asymptotic symmetry algebra \cite{Barnich:2006av} at null infinity with the classical central charges $c_L^\mathrm{flat}=0$ and $c^\mathrm{flat}_M=3/G_\mathrm{N}$, which corresponds to $\alpha^-=0$ and $\alpha^+=\frac{i}{\sqrt{2}G_\mathrm{N}}$ in our notation.

The results presented above are special cases ($N=2$) of the general free-field constructions for the Carrollian contractions of $W_N$-algebras that we explore in this article using the Miura transformation.

\section{Classical Carrollian $W_N$-algebras: free-field construction via Miura transformation}
\label{sec:Miura}
In this section, we first give a brief review of the free-field construction of classical $W_N$-algebras based on the (classical) Miura transformation. We then discuss the combinations of free fields and of higher-spin fields that need to be fixed in a Carrollian contraction. It is shown how such a contraction leads to a free-field construction of classical Carrollian $W_N$-algebras. We illustrate the result for the case $N=2$. 

\subsection{Classical $W_N$-algebras and Miura transformation}

A systematic construction of the $W_N$-algebra in terms of free fields is provided by the classical Miura transformation \cite{Dickey:1997wia}
\begin{equation}
    (\alpha \partial - J_1)(\alpha \partial - J_2) \cdots (\alpha \partial - J_N) = (\alpha\partial)^N - \sum_{k=2}^{N} u_k (\alpha\partial)^{N-k}\,,
\label{Miura-transform}
\end{equation}
where we follow conventions of~\cite{Fredenhagen:2024}.
This relation implicitly expresses the fields $u_i$ in terms of polynomials of spin-1 currents $J_j$ and their derivatives. We work on the cylinder and consider the relation at fixed time, so that the time variable can be neglected. The expansion of the free spin-1 fields $J_j$ is
\begin{equation}
    J_j(\theta)=i\sum_n J_{j,n}e^{in\theta}\,,
\end{equation}
with a $2\pi$-periodic angle-variable $\theta$. Derivatives with respect to $\theta$ are denoted simply by $\partial$. The currents can be regarded as derivatives of free fields, $J_j=i\partial \phi_j$. They are normalized such that their Poisson brackets read
\begin{equation}\label{PoissonbracketsJi}
    [J_{i,m},J_{j,n}]_\mathrm{cl} = \Big(\delta_{i,j}-\frac{1}{N}\Big) \delta_{m,-n}\,.
\end{equation}
Here we regard the rescaled Poisson brackets $[\_ ,\_ ]_\mathrm{cl}=i\{\_ ,\_ \}$ as `classical commutators', which makes comparison with the quantum case more convenient.
The fields $J_i$ are not all independent, but satisfy
\begin{equation}
    J_1+J_2+\cdots+J_N = 0\,,
\end{equation}
consistent with the brackets~\eqref{PoissonbracketsJi}.

The field $u_2$ can be identified with the energy-momentum tensor $T$. From the Miura transformation~\eqref{Miura-transform} one obtains
\begin{equation}
    T= u_2 = - \sum_{i<j} J_i J_j +\alpha \sum_{j>1} (j-1) \partial J_j\,.
\end{equation}
In the example $N=2$, one recovers the (classical version of) the free-field construction~\eqref{freefieldconofT} of $T$ in terms of $J=J_2=-J_1$. The fields $u_k$ are spin-$k$ fields and generate the classical $W_N$-algebra with (classical) central charge 
\begin{equation}
    c^{\mathrm{cl}} = -(N-1)N(N+1)\alpha^2\,.
    \label{classical-central-charge}
\end{equation}

\subsection{Preparing the Carrollian limit}

For the free-field realization of the Carrollian $W_N$-algebra that we are going to construct from the contraction of two copies of $W_N$-algebras, we first identify suitable combinations of the fields $u_k$ and $\bar{u}_k$ that are kept fixed during the contraction. For each copy, there exists a Miura transformation (with parameters $\alpha$ and $\bar{\alpha}$, respectively) that expresses the higher-spin currents in terms of free fields $J_j$ or $\bar{J}_j$. To analyze how the Miura transformation behaves in the Carrollian limit, we consider the combinations 
\begin{align}
    \mU_k^+ (\theta)&= \frac{\epsilon^k}{2}\big(u_k(\theta)+\bar{u}_k(\theta)\big)\,,& 
    \mU_k^-(\theta)&=\frac{\epsilon^{k-2}}{2}\big(u_k(\theta)-\bar{u}_k(\theta)\big)\,,
    \label{Carroll-spin-fields-def}
\end{align}
which are convenient for a Carrollian contraction. Similarly, we construct Carroll-compatible combinations of the spin-1 fields,
\begin{align}
         \mathcal{J}^+_i(\theta)&=\frac{\epsilon}{2}\big(J_i(\theta)+\bar{J}_i(\theta)\big)\,, &
         \mathcal{J}_i^-(\theta)&=\frac{\epsilon^{-1}}{2}\big(J_i(\theta)-\bar{J}_i(\theta)\big)\,.
    \label{Carrollian-fields}
\end{align}
Due to the appearance of the fields $J_i$ in the factors $(\alpha \partial-J_i)$ in the Miura transform~\eqref{Miura-transform}, it is natural to rescale the coefficients $\alpha$ and $\bar{\alpha}$ such that the combinations
\begin{align}
    \alpha^+ &= \frac{\epsilon}{2}\bigl(\alpha+\bar\alpha\bigr)\,,&
    \alpha^- &= \frac{\epsilon^{-1}}{2}\bigl(\alpha-\bar\alpha\bigr),
\end{align}
are kept fixed in the limit. For later convenience, we also state the inverted expressions,
\begin{subequations}\label{Carrolltransfinverted}
\begin{align}
    u_k(\theta) &= \epsilon^{-k}  \,\mU_k^+(\theta)+\epsilon^{2-k} \,\mU_k^-(\theta)\,, & \bar{u}_k(\theta) &= \epsilon^{-k} \, \mU^+_k(\theta)-\epsilon^{2-k}\, \mU_k^-(\theta)\,,\\
    J_i(\theta) &= \epsilon^{-1}  \,\mathcal{J}_i^+(\theta)+\epsilon \,\mathcal{J}_i^-(\theta)\,, & \bar{J}_i(\theta) &= \epsilon^{-1} \, \mathcal{J}^+_i(\theta)-\epsilon \,\mathcal{J}_i^-(\theta)\,,\label{J-epxressed-by-mJ}\\
    \alpha &= \epsilon^{-1} \alpha^+ + \epsilon \,\alpha^-\,, & \qquad \bar\alpha &= \epsilon^{-1}\alpha^+ - \epsilon \,\alpha^-\,.\label{alpha-rescaled}
\end{align}
\end{subequations}
Comparing the mode expansions of the Carroll-compatible fields 
\begin{equation}
    \mJ^{\pm}_j(\theta) = i\sum_n \mJ^{\pm}_{j,n}e^{in\theta}\,, 
\end{equation}
with those of the original fields
\begin{align}
    J_j (\theta) &= i\sum_n J_{j,n}e^{in\theta}\, ,& \bar{J}_j(\theta) &= i\sum_n \bar{J}_{j,n} e^{-in\theta}\,,
\end{align}
we get the following relation between the modes:
\begin{align}
        \mathcal{J}^{+}_{i,n}&=\frac{\epsilon}{2}\left(J_{i,n}+\bar{J}_{i,-n}\right)\, ,&
        \mathcal{J}^{-}_{i,n}&=\frac{\epsilon^{-1}}{2} \left(J_{i,n}-\bar{J}_{i,-n}\right)\,.
        \label{Carroll-generators}
\end{align}
Starting from the (classical) commutators
\begin{equation}
    \left[\mathcal{J}_{i,m},\mathcal{J}_{j,n}\right]_\mathrm{cl} = m\,\delta_{m,-n}\Big(\delta_{ij}-\frac{1}{N}\Big)\,,
\end{equation}
we find the commutators
\begin{align}
    \left[\mathcal{J}_m^{(i)+},\mathcal{J}_n^{(j)+}\right]_\mathrm{cl} &= 0\,,&
    \left[\mathcal{J}_m^{(i)-},\mathcal{J}_n^{(j)-}\right]_\mathrm{cl} &= 0\,,&
    \left[\mathcal{J}_m^{(i)-},\mathcal{J}_n^{(j)+}\right]_\mathrm{cl} &= \frac{m}{2}\delta_{m,-n}\Big(\delta^{ij}-\frac{1}{N}\Big)\,.
    \label{fundamental-free-field-commutators}
\end{align}

\subsection{Carrollian Miura transformation}

Expressing the fields $J_i,\, \bar J_i,\, u_k$ and $\bar u_k$ in terms of the Carroll-compatible fields $\mJ_i^\pm$, $\mU_k^\pm$, the Miura transformations become
\begin{multline}
    \underbrace{\Big(\frac{1}{\epsilon} (\alpha^+ \partial - \mJ_1^+) + \epsilon (\alpha^- \partial - \mJ_1^-)\Big) \cdots  \Big(\frac{1}{\epsilon} (\alpha^+ \partial - \mJ_N^+) + \epsilon (\alpha^- \partial - \mJ_N^-)\Big)}_{MT_L} \\= \underbrace{\big((\epsilon^{-1}\alpha^+ + \epsilon \alpha^-)\partial\big)^N - \sum_{k=2}^{N} \big(\epsilon^{-k}\mU^+_k + \epsilon^{2-k}\mU^-_k\big) \big((\epsilon^{-1}\alpha^+ + \epsilon \alpha^-)\partial\big)^{N-k}}_{MT_R}\,,
\label{MLMR}
\end{multline}
and
\begin{multline}
    \underbrace{\Big(\frac{1}{\epsilon} (\alpha^+ \partial - \mJ_1^+) - \epsilon (\alpha^- \partial - \mJ_1^-)\Big) \cdots  \Big(\frac{1}{\epsilon} (\alpha^+ \partial - \mJ_N^+) - \epsilon (\alpha^- \partial - \mJ_N^-)\Big)}_{\overline{MT}_L} \\= \underbrace{\big((\epsilon^{-1}\alpha^+ - \epsilon \alpha^-)\partial\big)^N - \sum_{k=2}^{N} \big(\epsilon^{-k}\mU^+_k - \epsilon^{2-k}\mU^-_k\big) \big((\epsilon^{-1}\alpha^+ - \epsilon \alpha^-)\partial\big)^{N-k}}_{\overline{MT}_R}\,,
\label{barMLbarMR}
\end{multline}
where we introduced notations for the left- and right-hand side of the Miura transformations. We now treat the Miura transformation as similar to a spin-$N$ field and form the combinations
\begin{align}
    \mathcal{MT}_{L/R}^+ &= \frac{\epsilon^N}{2} \big(MT_{L/R} + \overline{MT}_{L/R}\big) \,,&
    \mathcal{MT}_{L/R}^- &= \frac{\epsilon^{N-2}}{2} \big(MT_{L/R} - \overline{MT}_{L/R}\big) \,.
\end{align}
We work out the corresponding expressions below. Starting with $\mathcal{MT}_R^+$, we find
\begin{equation}
    \mathcal{MT}_R^+ \xrightarrow{\epsilon\to 0} (\alpha^+ \partial)^N - \sum_{k=2}^N \mU^+_k (\alpha^+ \partial)^{N-k}\,.
\end{equation}
Similarly one obtains
\begin{equation}
    \mathcal{MT}_L^+ \xrightarrow{\epsilon\to 0}
    (\alpha^+ \partial - \mJ^+_1) \cdots (\alpha^+ \partial -\mJ^+_N)\,.
\end{equation}
For the other combination, one gets
\begin{equation}
    \mathcal{MT}_R^- \xrightarrow{\epsilon\to 0} N (\alpha^+)^{N-1} \alpha^- \partial^N -\sum_{k=2}^N \big( (N-k) \mU^+_k (\alpha^+)^{N-k-1} \alpha^- + \mU^-_k (\alpha^+)^{N-k}  \big)\partial^{N-k}\,,
\end{equation}
and
\begin{equation}
    \mathcal{MT}_L^- \xrightarrow{\epsilon\to 0} \sum_{j=1}^N (\alpha^+\partial -\mJ_1^+) \cdots (\alpha^+\partial - \mJ_{j-1}^+)(\alpha^- \partial - \mJ_j^-)(\alpha^+\partial - \mJ_{j+1}^+)\cdots (\alpha^+\partial - \mJ^+_N)\,.
\end{equation}
In total, we arrive at the following result, which implicitly expresses the higher-spin fields of the Carrollian $W_N$-algebra in terms of the free fields $\mJ^\pm_i$:
\begin{align}\label{CarrollMT1}
    (\alpha^+ \partial - \mJ^+_1) \cdots (\alpha^+ \partial -\mJ^+_N) &= (\alpha^+ \partial)^N - \sum_{k=2}^N \mU^+_k(\alpha^+ \partial)^{N-k}\,,\\
    \sum_{j=1}^N (\alpha^+\partial -\mJ_1^+) \cdots (\alpha^-\partial -\mJ_j^-) \cdots (\alpha^+\partial - \mJ^+_N) &= N(\alpha^+)^{N-1}\alpha^-\partial^N -\sum_{k=2}^N \mU^-_k (\alpha^+\partial )^{N-k}\nonumber\\
    &\quad-\sum_{k=2}^N (N-k)\mU^+_k(\alpha^+)^{N-k-1}\alpha^- \partial^{N-k}\,.
    \label{CarrollMT2}
\end{align}
The first equation~\eqref{CarrollMT1} can be used to define the $\mU_i^+$, the second equation~\eqref{CarrollMT2} then encodes the $\mU^-_i$. There is also an alternative way to present these relations. When we take the derivative of the first equation~\eqref{CarrollMT1} with respect to $\alpha^+$, we find
\begin{multline}
    \sum_{j=1}^N(\alpha^+ \partial - \mJ^+_1) \cdots (\alpha^+ \partial - \mJ^+_{j-1})\partial (\alpha^+ \partial - \mJ^+_{j+1})\cdots (\alpha^+ \partial -\mJ^+_N) \\= N(\alpha^+)^{N-1} \partial^N - \sum_{k=2}^N \big(\mU^+_k (N-k)(\alpha^+)^{N-k-1} +  (\alpha^+)^{N-k}\partial_{\alpha^+}\mU^+_k\big)\partial^{N-k}\,,
\end{multline}
where we take into account that the fields $\mU^+_j$ expressed in terms of the $\mJ^+_i$ also depend on $\alpha^+$. Combining this with~\eqref{CarrollMT2}, one can derive the relation
\begin{equation}\label{CarrollMT2a}
    \sum_{j=1}^N (\alpha^+\partial -\mJ_1^+) \cdots \mJ_j^- \cdots (\alpha^+\partial - \mJ^+_N) = \sum_{k=2}^N \big(\mU^-_k -\alpha^- \partial_{\alpha^+}\mU^+_k\big) (\alpha^+\partial )^{N-k} \,.
\end{equation}
Equations~\eqref{CarrollMT1} together with~\eqref{CarrollMT2} or~\eqref{CarrollMT2a} constitute a Carrollian Miura transformation which implicitly expresses the fields $\mU^\pm_k$ of a Carrollian $W_N$-algebra in terms of free fields satisfying
\begin{equation}
    \sum_i \mJ_i^\pm = 0 \, .
\end{equation}
Among the fields, we find the contractions arising from the components of the energy-momentum tensor, $T=u_2$ and $\bar{T}=\bar{u}_2$. Indeed, the field $\mU^-_2$ coincides up to a factor of $2$ with the field $\mathcal{T}$ whose modes are the Carrollian generators $\mL_n$,
\begin{equation}
    \mathcal{T}(\theta) =\sum_n \mL_n e^{in\theta} = \sum_n \big(L_n-\bar{L}_{-n}\big)e^{in\theta} = T(\theta)-\bar{T}(\theta)= 2\,\mU^-_2(\theta) \,,
\end{equation}
and similarly $\mU_2^+$ corresponds to the field $\mM$ that contains the modes $\mM_n$,
\begin{equation}
    \mathcal{M}(\theta) = \sum_n \mM_n e^{in\theta} = \epsilon^2 \sum_n \big(L_n + \bar{L}_{-n}\big)e^{in\theta} = \epsilon^2 \big(T(\theta)+\bar{T}(\theta)\big) = 2\,\mU_2^+(\theta)\,.
\end{equation}
Thus, the Carrollian $W_N$-algebras contain the Carrollian conformal algebra. The (classical) central charges can be inferred from the central charges $c^{\mathrm{cl}}$ and $\bar{c}^{\mathrm{cl}}$ of the $W_N$-algebras given in~\eqref{classical-central-charge}, and one finds
\begin{align}\label{classical-central-charge-Carroll-L}
    c^{\mathrm{cl}}_L = c^{\mathrm{cl}}-\bar{c}^{\mathrm{cl}} = -(N-1)N(N+1) \big(\alpha^2 -\bar{\alpha}^2\big) = -4(N-1)N(N+1)\alpha^+ \alpha^-\,,
\end{align}
and 
\begin{equation}\label{classical-central-charge-Carroll-M}
    c^{\mathrm{cl}}_M = \lim_{\epsilon\to 0} \epsilon^2 \big(c^{\mathrm{cl}}+\bar{c}^{\mathrm{cl}}\big) = -2 (N-1)N(N+1)(\alpha^+)^2\,. 
\end{equation}

\subsection{The example $N=2$}
For $N=2$, there are only the two fields $\mU^\pm_2$. From~\eqref{CarrollMT1} we read off
\begin{equation}\label{mM-from-free-fields}
    \mM= 2\,\mU^+_2 = 2\,\mJ^+ \mJ^+ + 2\alpha^+ \partial \mJ^+ 
\end{equation}
where we define $\mJ^+=\mJ^+_2 = -\mJ_1^+$.
From~\eqref{CarrollMT2a} we find
\begin{align}
    \mT= 2\, \mU_2^- &= 4\,\mJ^+ \mJ^- +2\alpha^+ \partial \mJ^- + \alpha^-\partial_{\alpha^+}\mU_2^+ \\
    &= 4\,\mJ^+ \mJ^- +2\alpha^+ \partial \mJ^- +2\alpha^-\partial \mJ^+\,.\label{mT-from-free-fields}
\end{align}
We use this example to illustrate how these expressions emerge in the $\epsilon\to 0$ limit. We start from the field construction of the Virasoro algebras
\begin{align}
    T &= J\,J + \alpha\,  \partial J\,, &
    \bar{T} &= \bar{J}\,\bar{J} + \bar{\alpha}\,\partial\bar{J}\, .
\end{align}
Expressing $J$ and $\bar{J}$ in terms of $\mJ^\pm$ (see~\eqref{J-epxressed-by-mJ}) we obtain for $\mM=2\,\mU_2^+$
\begin{align}
    \mM&=\epsilon^2 \big( T + \bar{T}\big)\\
    &=\epsilon^2 \big( (\epsilon^{-1}\mJ^+ + \epsilon \mJ^-)^2 + (\epsilon^{-1}\alpha^+ +\epsilon \alpha^-\big)\partial (\epsilon^{-1}\mJ^+ + \epsilon \mJ^-))\nonumber\\
    &\quad + \epsilon^2 \big( (\epsilon^{-1}\mJ^+ - \epsilon \mJ^-)^2 + (\epsilon^{-1}\alpha^+ -\epsilon \alpha^-\big)\partial (\epsilon^{-1}\mJ^+ - \epsilon \mJ^-))\\
    & \xrightarrow{\epsilon\to 0} 2 \mJ^+\mJ^++ 2\alpha^+\,\partial \mJ^+\,.
    \label{CarrollM-N=2}
\end{align}
Similarly, for $\mT=2\mU^-_2$ we find 
\begin{align}
    \mT&= T-\bar{T}\\
    &= \big( (\epsilon^{-1}\mJ^+ + \epsilon \mJ^-)^2 + (\epsilon^{-1}\alpha^+ +\epsilon \alpha^-\big)\partial (\epsilon^{-1}\mJ^+ + \epsilon \mJ^-))\nonumber\\
    &\quad - \big( (\epsilon^{-1}\mJ^+ - \epsilon \mJ^-)^2 + (\epsilon^{-1}\alpha^+ -\epsilon \alpha^-\big)\partial (\epsilon^{-1}\mJ^+ - \epsilon \mJ^-))\\
    &= 4\,\mJ^+\mJ^- + 2\alpha^+\,\partial \mJ^- + 2\alpha^-\,\partial \mJ^+\,.
    \label{CarrollT-N=2}
\end{align}
As expected, in the limit $\epsilon\to 0$, the free-field construction of the Virasoro algebras reduces to the free-field construction of $\mM$ and $\mT$ (see~\eqref{mM-from-free-fields} and~\eqref{mT-from-free-fields} above) that is encoded in the Carrollian Miura transformation. Note that when the expressions~\eqref{mM-from-free-fields} and~\eqref{mT-from-free-fields} are expanded in modes on the plane, they lead to the explicit expressions given in~\eqref{FreefieldconofCCA}.
\section{Quantum Carrollian $W_N$-algebras}
\label{sec:quantum}

In this section, we study quantum counterparts of the classical free-field constructions given above. We consider a Carrollian contraction of the quantum $W_N$-algebras in two different setups. First, we consider a \emph{flipped} construction where we reverse the time direction and hence the normal ordering prescription for one of the copies of the relativistic $W_N$-algebras from which we start. The resulting algebras are built from free fields with usual normal ordering prescriptions, and these algebras are isomorphic to the quantum Galilean $W_N$-algebras. Second, we consider a \emph{symmetric} construction, in which time and the normal ordering prescription are kept in both copies of the $W_N$-algebra. The resulting quantum Carroll $W_N$-algebras can be formulated in terms of free fields using an averaged ordering prescription. The basic commutation relations turn out to coincide with the ones of the classical $W_N$-algebras.

\subsection{Quantum Miura transformation and normal ordering}

To obtain quantum $W_N$-algebras, one can again use the Miura transformation~\eqref{Miura-transform}, where the fields $J_j$ and $u_k$ are considered to be quantum fields \cite{Fateev:1987zh}. The products appearing on the left-hand side are now given as normal-ordered products in the usual sense in a two-dimensional Euclidean conformal field theory. Normal ordering depends crucially on the identification of Euclidean time,
\begin{equation}
    :\!\mO _1(z) \mO_2(w)\!: \,= \mathcal{T} \big( \mO_1(z) \mO_2(w)\big) - \text{singular part}\,,
\end{equation}
where $\mathcal{T}$ denotes the (Euclidean) time ordering, and we subtract the singular part in an expansion in powers of $(z-w)$. For free spin-1 fields on the plane in radial quantization, 
this normal ordering is simply mode normal ordering, where modes with positive mode numbers are ordered to the right. If one reverses the direction of time on the plane, normal ordering for free fields becomes \emph{mode anti-normal ordering}, where modes with positive mode numbers are ordered to the left.

\subsection{Carrollian contractions in conformal field theories}

To motivate the approach for the flipped construction of quantum Carrollian $W_N$-algebras, we first consider the contraction of two copies of the Virasoro algebra to the Carrollian conformal algebra~\eqref{CCA}. On the Euclidean cylinder, the components of the energy-momentum tensor can be expanded as
\begin{align}
    T(\tau,\theta)&=\sum_n L_n e^{-n\tau+in\theta}\,, &
    \bar{T}(\tau,\theta)&=\sum_n \bar L_n e^{-n\tau-in\theta}\,.
\end{align}
A Carrollian limit can be obtained by rescaling the time $\tau\to\epsilon^2\tau$ while keeping the space variable $\theta$ unchanged. When expanding the fields $T$ and $\bar{T}$ for small $\epsilon$ and introducing the Carrollian generators $\mL_m$ and $\mM_n$ as in~\eqref{Carroll-Virasoro-trafo}, one can build the combinations
\begin{align}
    \mathcal{T}_C(\tau,\theta)&=T(\epsilon^2 \tau,\theta)-\bar{T}(\epsilon^2 \tau,\theta)  \\ 
    &=\sum_n \left[(L_n-\bar{L}_{-n}) - n\epsilon^2\tau (L_n+\bar{L}_{-n})\right] e^{in\theta} + \mathcal{O}(\epsilon^4) \\
    & \xrightarrow{\epsilon \to 0}\sum_n \left[\mathcal{L}_n - n\tau \mathcal{M}_n \right] e^{in\theta} \,,\\
    \mathcal{M}_C(\tau,\theta)&=\epsilon^2(T(\epsilon^2 \tau,\theta)+\bar{T}(\epsilon^2 \tau,\theta))  \\ &=\epsilon^2\bigg(\sum_n \left[(L_n+\bar{L}_{-n}) - n\epsilon^2\tau (L_n-\bar{L}_{-n})\right] e^{in\theta} + \mathcal{O}(\epsilon^4)\bigg) \\&\xrightarrow{\epsilon \to 0} \sum_n \mathcal{M}_n e^{in\theta} \,.
\end{align}
The modes $\mL_m$ and $\mM_n$ appearing in the expansion of the Carrollian fields $\mathcal{T}_C$ and $\mM_C$ satisfy the Carrollian conformal algebra~\eqref{CCA}. 

The same algebra can be obtained without rescaling time and space, by considering instead combinations of the fields $T$ and $\bar{T}$ where we reverse the time direction for $\bar{T}$,
\begin{subequations}\label{FieldContractionCarroll}
\begin{alignat}{2}
    \tilde{\mathcal{T}}(\tau,\theta)&=T(\tau,\theta)-\bar{T}(-\tau,\theta) &&\xrightarrow{\epsilon\to0} \sum_n \tilde{\mathcal{L}}_n e^{-n\tau+in\theta}\,, \\
    \tilde{\mathcal{M}}(\tau,\theta)&=\epsilon^2(T(\tau,\theta)+\bar{T}(-\tau,\theta)) &&\xrightarrow{\epsilon\to0} \sum_n \tilde{\mathcal{M}}_n e^{-n\tau+in\theta}\,.
\end{alignat}
\end{subequations}
Although no Carrollian limit of the space and time coordinates was involved, one obtains fields whose modes $\mL_m$ and $\mM_n$ satisfy the Carrollian conformal algebra~\eqref{CCA}.

\subsection{The flipped construction: a Carrollian road to quantum Galilean $W_N$-algebras}\label{sec:flipped}

We follow the strategy outlined in the previous subsection to obtain quantum counterparts of Carrollian $W_N$-algebras. For this purpose, we define the combinations
\begin{align}
    \tilde{\mU}_k^+(\tau,\theta) &= \frac{\epsilon^k}{2}\big(u_k(\tau,\theta)+\bar{u}_k(-\tau,\theta)\big)\,,&
    \tilde{\mU}_k^-(\tau,\theta) &=\frac{\epsilon^{k-2}}{2}\big(u_k(\tau,\theta)-\bar{u}_k(-\tau,\theta)\big)\,,
\end{align}
as well as
\begin{align}
         \mathcal{J}^+_i(\tau,\theta)&=\frac{\epsilon}{2}\big(J_i(\tau,\theta)+\bar{J}_i(-\tau,\theta)\big)\,,&
         \mathcal{J}_i^-(\tau,\theta)&=\frac{\epsilon^{-1}}{2}\big(J_i(\tau,\theta)-\bar{J}_i(-\tau,\theta)\big)\,,
\end{align}
similarly to what we did in~\eqref{Carroll-spin-fields-def} and~\eqref{Carrollian-fields} for the classical fields. Normal ordering for $\mJ^\pm_i$ corresponds to normal ordering for the fields $J_i$ and time-reversed normal ordering for $\bar{J}_i$. We call this the \emph{flipped construction}.

In this flipped setup, we can follow precisely the same steps as in the classical case to arrive at a quantum version of the Carrollian Miura transformation\footnote{Note that normal ordering is not needed in the definition of $\mU^+_k$ as the $\mJ^+_i$ are all commuting.}
\begin{align}\label{CarrollQMT}
    (\alpha^+ \partial - \mJ^+_1) \cdots (\alpha^+ \partial -\mJ^+_N) &= (\alpha^+ \partial)^N - \sum_{k=2}^N \tilde{\mU}^+_k (\alpha^+ \partial)^{N-k}\,,\\
    \sum_{j=1}^N :\!(\alpha^+\partial -\mJ_1^+) \cdots \mJ_j^- \cdots (\alpha^+\partial - \mJ^+_N)\!: &= \sum_{k=2}^N \big(\tilde{\mU}^-_k -\alpha^- \partial_{\alpha^+}\tilde{\mU}^+_k\big) (\alpha^+\partial )^{N-k} \,.
\end{align}
These expressions define a free-field realization of a quantum algebra whose classical version is the Carrollian $W_N$-algebra. Due to the flip in the time direction, this algebra should rather be viewed as the quantum Galilean $W_N$-algebra. We discuss the relation to a Galilean contraction of the Miura transformation in Appendix~\ref{app:Galilean}.

The central charges $\tilde{c}_L$ and $\tilde{c}_M$ can be obtained as limits of the corresponding expressions built from the central charges $c$ and $\bar{c}$ of the quantum $W_N$-algebras. They differ from the classical expression~\eqref{classical-central-charge} by a correction coming from normal ordering, and they are given by (see for example~\cite{Fredenhagen:2024}),
\begin{align}
    c & = (N-1) \big( 1-N(N+1)\alpha^2\big)\,,\\
    \bar{c} &= (N-1) \big( -1-N(N+1)\bar{\alpha}^2\big)\,.
    \label{cbar}
\end{align}
Note that the sign of the quantum correction of the central charge $\bar{c}$ differs from that of $c$ due to the reversed normal ordering (anti-normal ordering, see Appendix~\ref{B}). The central charges of the Carrollian conformal algebra are then given by
\begin{align}
    \label{cL}
    \tilde{c}_L = c-\bar{c} = (N-1)\big(2-4 N(N+1)\alpha^+ \alpha^-\big)\,,
\end{align}
and 
\begin{equation}
    \label{cM}
    \tilde{c}_M = \lim_{\epsilon\to 0} \epsilon^2 \big(c+\bar{c}\big) = -2 (N-1)N(N+1)(\alpha^+)^2\,. 
\end{equation}
This matches the expressions for $N=2$ presented in~\eqref{Carroll-central-charges-N=2}. One observes that for $\alpha^\pm=0$, these expressions coincide with the ones obtained for the quantization of $D=N-1$ free fields in the `flipped vacuum' \cite{Bagchi:2020fpr}.

As an example, we provide the expressions for the spin-2 and spin-3 fields. From~\eqref{CarrollQMT} we can read off 
\begin{align}
    \tilde{\mU}^+_2 &= -\sum_{i<j} \mJ_i^+ \mJ_j^+ + \sum_{j} (j-1) \alpha^+\partial \mJ_j^+\,,\\
    \tilde{\mU}^-_2 &= -\sum_{i<j} \big( :\!\mJ_i^- \mJ_j^+ \!:+ :\!\mJ_i^+ \mJ_j^-\!: \big) + \sum_j (j-1) \big( \alpha^- \partial \mJ_j^+ + \alpha^+ \partial \mJ_j^-\big)\,,\displaybreak[0]\\
    \tilde{\mU}^+_3 &= \sum_{i<j<k} \mJ_i^+ \mJ_j^+ \mJ_k^+ - \sum_{i<j} \big((i-1) \alpha^+ \partial \mJ_i^+ \,\mJ_j^+ + (j-2) \alpha^+ \mJ_i^+\,\partial \mJ_j^+\big) \nonumber\\
    &\quad+ \sum_{j\geq 3} \binom{j-1}{2} (\alpha^+)^2 \partial^2 \mJ_j^+\,,\displaybreak[0]\\
    \tilde{\mU}^-_3 &= \sum_{i<j<k} \big(:\!\mJ_i^- \mJ_j^+ \mJ_k^+\!: +:\! \mJ_i^+ \mJ_j^- \mJ_k^+\!: + :\! \mJ_i^+ \mJ_j^+ \mJ_k^- \!: \big)\nonumber \\
    &\quad - \sum_{i<j} (i-1) \big(\alpha^- :\!\partial \mJ_i^+ \,\mJ_j^+\!: +\alpha^+ :\!\partial \mJ_i^- \,\mJ_j^+ \!: +\alpha^+ :\!\partial \mJ_i^+ \,\mJ_j^-\!:\big) \nonumber\\
    &\quad -\sum_{i<j} (j-2) \big(\alpha^- :\!\mJ_i^+\,\partial \mJ_j^+ \!:+ \alpha^+ :\!\mJ_i^-\,\partial \mJ_j^+ \!:+ \alpha^+ :\!\mJ_i^+\,\partial \mJ_j^-\!: \big) \nonumber\\
    &\quad + \sum_{j\geq 3} \binom{j-1}{2} \big( 2\alpha^+ \alpha^- \partial^2 \mJ_j^+ + (\alpha^+)^2 \partial^2 \mJ_j^-\big) \,.
\end{align}
Up to a factor of $2$ we identify $\tilde{\mU}_2^\pm$ with the fields $\tilde{\mT}$ and $\tilde{\mM}$ that contain the modes $\tilde{\mL}_m$ and $\tilde{\mM}_m$,
\begin{align}
    \tilde{\mT} &= 2 \,\tilde{\mU}^-\,, & \tilde{\mM} &= 2\,\tilde{\mU}^+\,.
\end{align}
We work out the example of $N=3$ explicitly in Appendix~\ref{sec:W3}.

We have arrived at the expression~\eqref{CarrollQMT} for the (flipped) quantum counterpart of the Carrollian Miura transformation by considering fields on the (Euclidean) cylinder. We could also have done this on the plane, but one has to be aware that reversing radial time, $\bar{z}\to 1/z$, induces non-trivial transformations on the fields (in particular because the fields $u_k$ are not quasi-primary). The final expression~\eqref{CarrollQMT}, however, can again be evaluated on the plane.

\subsection{Proper quantum Carrollian $W_N$-algebras}

In the flipped construction above, we reversed the time direction in one sector, thereby not constructing the Carrollian contraction of two copies of the same $W_N$-algebra, but instead starting from one copy of the $W_N$-algebra and one copy of an anti-normal-ordered $W_N$-algebra. We now consider a symmetric construction where we do not reverse the time direction (and correspondingly the normal ordering) in one copy.

Let us briefly illustrate the difference in the case $N=2$, which leads us back to the discussion of Subsection~\ref{sec:freefieldconstruction}. Although the Galilean and Carrollian quantum algebras are in this case isomorphic, the corresponding free-field constructions differ. We start from the usual Sugawara-type 
free-field construction of the Virasoro algebra at $c=1$ where\footnote{This corresponds to the case $N=2$ and $\alpha=0$ of the free-field construction via the Miura transformation.} $T=:\!JJ\!:$. Normal ordering is only relevant for the zero mode,
\begin{equation}
L_0 = \sum_m :\!J_m J_{-m}\!: \,= 2\sum_{m>0} J_{-m}J_m + J_0J_0\,.  \end{equation}
$\bar{L}_0$ is constructed similarly from $\bar{J}_m$ where we now use the same normal ordering as for $L_0$. We then rewrite
\begin{align}
    J_m &= \epsilon^{-1}\mJ^+_m + \epsilon \,\mJ^-_{m} \,,& \bar{J}_m&= \epsilon^{-1} \mJ^+_{-m} - \epsilon \,\mJ^-_{-m}\,.
\end{align}
The corresponding Carrollian mode $\mL_0$ can be written via averaged ordering (see~\eqref{symmordering}):
\begin{align}
    \mL_0 &= L_0 -\bar{L}_0\\
    &= 2\sum_{m>0} J_{-m}J_m + J_0 J_0 - 2\sum_{m>0} \bar{J}_{-m}\bar{J}_m -\bar{J}_0\bar{J}_0\\
    &= 2\sum_{m>0} \big(\mJ^+_{-m}\mJ^-_m + \mJ^-_{-m}\mJ^+_m\big)  + 2\sum_{m>0}\big(\mJ^+_m \mJ^-_{-m} + \mJ^-_m\mJ^+_{-m}\big) + 4\,\mJ^+_0\mJ^-_0\\
    &=4 \sum_m \triplenormord{\mJ^+_{m}\mJ^-_{-m}}\,.
\end{align}
With this ordering prescription, it is straightforward to check that the modes $\mL_n$ satisfy the Virasoro algebra with central charge $c_L=0$. 

Carrying out the full construction for $N=2$ including the linear terms in the currents with coefficients $\alpha$ and $\bar{\alpha}$ (rescaled as before, see~\eqref{alpha-rescaled}), we arrive at the free-field construction in~\eqref{2ndFreefieldconofCCA} with the central charges~\eqref{Carroll-central-charges-N=2-classical}. The symmetric Carrollian contraction thus leads to the same algebraic structure as the flipped one, but with shifted central charges compared to~\eqref{Carroll-central-charges-N=2}.

For general $N$, the construction again yields the same expressions~\eqref{CarrollQMT} for the higher-spin fields $\mU^\pm$, except that the normal ordering of the currents $\mJ^\pm_i$ is replaced by the averaged ordering $\triplenormord{\ }$. This is because the terms involving $\bar J_{i,-n}$ are mode normal-ordered in the index $-n$, which places the operators $\mathcal J^{\pm}_n$ in anti-normal order with respect to $n$. Combining these contributions with the usual normal-ordered terms from $J_{i,n}$ leads precisely to the averaged prescription.

The fields $\mU^+_j$ are built only from $\mJ^+_i$, which all commute and do not require an ordering prescription. The fields $\mU^-_j$ are built as sums of products of $\mJ^+$ fields and at most a single $\mJ^-$ fields. Schematically, a contribution of this kind can be written as
\begin{equation}
    \triplenormord{\mJ^-_{j,n} \mJ^+_{[<n]} \mJ^+_{[\geq n]}} = \frac{1}{2} \Big(  \mJ^+_{[<n]}\mJ^-_{j,n} \mJ^+_{[\geq n]} + \mJ^+_{[\geq n]}\mJ^-_{j,n} \mJ^+_{[<n]} \Big)\,,
\end{equation}
where $\mJ^+_{[\geq n]}$ represents a product of (commuting) modes $\mJ^+_{i,n_i}$ with mode numbers $n_i\geq n$, and $\mJ^-_{[<n]}$ similarly represents a product of modes $\mJ^+_{i,n_i}$ with $n_i<n$. Because of 
\begin{equation}\label{+-+commutator}
    \big[\mJ^+_{[<n]}, \big[\mJ^-_{j,n},\mJ^+_{[\geq n]}\big]\big] = 0\,,
\end{equation}
we can rewrite 
\begin{equation}
    \triplenormord{\mJ^-_{j,n} \mJ^+_{[<n]} \mJ^+_{[\geq n]}} 
    = \frac{1}{2} \Big(\mJ^-_{j,n} \mJ^+_{[<n]} \mJ^+_{[\geq n]}  + \mJ^+_{[<n]} \mJ^+_{[\geq n]}  \mJ^-_{j,n}  \Big) \,.
\end{equation}
For products that only contain one $\mJ^-$ mode, the averaged ordering coincides with the prescription to order the $\mJ^-$ once to the left and once to the right in an averaged combination.

With such an ordering prescription the basic commutation relations turn out to be the same as in the classical theory. Let us consider the possible commutators. Commutators of two $\mU^+$ fields vanish, and commutators of $\mU^-$ and $\mU^+$ only lead to terms involving the $\mU^+$ fields which commute and do not require any ordering prescription. Now consider the commutator of two fields of type $\mU^-$. We analyse a typical commutator that appears in such a computation: by only using the derivation property (which holds both in the classical and in the quantum case) we see that\footnote{Here, $\mJ^-_{1}$ stands for any mode $\mJ^-_{j_1,n_1}$ (and analogously $\mJ^-_2$) whereas $\mJ^+_{1,2}$ stands for any product of modes of the $+$-currents $\mJ^+$.}
\begin{align}
    & \big[ \mJ^-_1 \mJ^+_1 + \mJ^+_1 \mJ^-_1 , \mJ^-_2 \mJ^+_2 + \mJ^+_2 \mJ^-_2 \big]\nonumber \\
    &\quad = \mJ^-_1 \big[ \mJ^+_1,\mJ^-_2 \big] \mJ^+_2 + \mJ^-_1 \mJ^+_2 \big[\mJ^+_1,\mJ^-_2\big]  +\mJ^-_2 \big[\mJ^-_1,\mJ^+_2\big] \mJ^+_1 + \textcolor{blue}{\big[\mJ^-_1,\mJ^+_2 \big] \mJ^-_2 \mJ^+_1 }\nonumber\\
    & \qquad + \textcolor{blue}{\mJ^+_1 \mJ^-_2 \big[\mJ^-_1 , \mJ^+_2 \big] } + \mJ^+_1 \big[\mJ^-_1 , \mJ^+_2 \big] \mJ^-_2+ \big[\mJ^+_1 ,\mJ^-_2\big] \mJ^+_2 \mJ^-_1 + \mJ^+_2 \big[\mJ^+_1, \mJ^-_2 \big] \mJ^-_1 \,. \label{typicalcommutator}
\end{align}
As commutators of the form $\big[ \mJ^+_1 ,\mJ^-_2\big]$ only contain $+$-currents, the right hand side consists again of terms that do contain at most one $\mJ^-$. The fourth and fifth terms (denoted in blue) are the only ones where the $\mJ^-$ field is not at the left or at the right end of the product, but one can directly verify\footnote{Using that $\big[ \big[ \mJ^-_1 ,\mJ^+_2 \big],\big[ \mJ^-_2, \mJ^+_1 \big]\big]=0$.} that
\begin{equation}
    \textcolor{blue}{\big[\mJ^-_1,\mJ^+_2 \big] \mJ^-_2 \mJ^+_1 + \mJ^+_1 \mJ^-_2 \big[\mJ^-_1 , \mJ^+_2 \big]} = \big[ \mJ^-_1,\mJ^+_2 \big] \mJ^+_1 \mJ^-_2 + \mJ^-_2 \mJ^+_1 \big[ \mJ^-_1 ,\mJ^+_2 \big] \,.
\end{equation}
This allows us to rewrite~\eqref{typicalcommutator} as
\begin{align}
    & \big[ \mJ^-_1 \mJ^+_1 + \mJ^+_1 \mJ^-_1 , \mJ^-_2 \mJ^+_2 + \mJ^+_2 \mJ^-_2 \big]= 4 \,\Big(\triplenormord{\mJ^-_1 \big[ \mJ^+_1,\mJ^-_2 \big] \mJ^+_2 +\mJ^-_2 \big[\mJ^-_1,\mJ^+_2\big] \mJ^+_1 } \Big)\,.
    \label{typicalcommutator-result}
\end{align}
The outcome of the commutator~\eqref{typicalcommutator} is already in the averaged ordered form without the need to reorder, and hence the fields satisfy the commutation relations with the same structure constants as in the classical Carrollian $W_N$-algebra.

It might come as a surprise that the structure constants do not receive any quantum corrections due to the quantum Jacobi identity. The Jacobi identity involving two or three $\mU^+$ modes is trivially satisfied. For two $\mU^-$ and one $\mU^+$ mode, the result consists only of commuting $+$-modes. Hence the classical Jacobi identity implies the quantum one here. The only term that could lead to corrections comes from the Jacobi identity involving three $\mU^-$ modes. As the $\mU^-$ modes as well as the commutators of two $\mU^-$ modes are in the averaged ordering, a typical term in the double commutator is again of the form~\eqref{typicalcommutator-result}, and hence all such terms appear already in the averaged ordering. There is no need to reorder, and the classical Jacobi identity implies the quantum one.

The argument above also applies if one considers another basis (for example the basis that would arise from a contraction of the $W_N$-algebras in the commonly used $W$-basis where the higher-spin currents are primary fields) as long as the basis elements contain at most one current $\mJ^-$. Or, more generally, the argument carries over to other $W$-algebras than $W_N$: the argument above only relies on the fact that there are two types of basis elements, $W^\pm$, such that the $W^+$ terms commute, the commutator of two $W^-$ contains products involving at most one $W^-$, and the commutator of $W^+$ and $W^-$ is built solely from $W^+$ (and thus it commutes with any other $W^+$, similarly to~\eqref{+-+commutator}). 

\section{Discussion and outlook}
\label{sec:discussion}

In this work, we have constructed Carrollian $W_N$-algebras by performing an ultra-relativistic contraction of two copies of the standard $W_N$-algebra. Our approach uses a free-field realization based on the Miura transformation, yielding a systematic method to derive classical and quantum forms of the contracted algebra. We have discussed two quantum algebras that generalize the classical Carrollian $W_N$-algebra to commutator algebras: the quantum Carrollian $W_N$-algebra and the quantum Galilean $W_N$-algebra. The proper Carrollian $W_N$-algebra arises from a symmetric contraction which leads to an averaged ordering prescription for the non-linear terms. As we have discussed, the structure constants for the commutators of the basic generators coincide with those of the classical theory.

On the other hand, the Galilean $W_N$-algebra can be obtained either by a Galilean contraction or by a flipped Carrollian contraction, where the time direction is reversed in one sector. As a concrete example, we worked out the full structure of the Galilean $W_3$-algebra using the free-field construction (see Appendix~\ref{sec:W3}). 
Whereas specific examples of free-field constructions for the Galilean conformal and the Galilean $W_3$-algebra have appeared in the literature \cite{Gonzalez:2014tba,Banerjee:2015kcx,Ammon:2017vwt,Radobolja:2021nae}, we provide a novel, systematic free-field construction for general Carrollian and Galilean $W_N$-algebras, which also clarifies the subtle differences between the Galilean and the Carrollian case in the quantum regime. 

Our construction provides a framework for systematically extending Carrollian conformal symmetry, in close analogy with the role played by $W_N$-algebras in relativistic two-dimensional conformal field theory. In fact, Carrollian $W_N$-algebras appear naturally as asymptotic symmetries in flat space higher-spin gravity~\cite{Afshar:2013vka,Gonzalez:2013oaa}, which further motivates their study. Moreover, it is reasonable to expect that their free-field construction can also be recovered from a holographic perspective in a suitable gauge, in close parallel with the AdS case, where the free-field (Miura) realization of $W_N$-algebras emerges naturally in the diagonal gauge when analyzing asymptotic symmetries in three-dimensional higher-spin gravity~\cite{Campoleoni:2017xyl,Ojeda:2020bgz}.

It would be interesting to explore the possibility of defining a boundary Carrollian $W_N$-algebra from a single copy of the relativistic $W_N$-algebra, in analogy with recent work on boundary Carrollian CFTs~\cite{Bagchi:2024qsb,Bagchi:2025jgu}. In this context, a contraction of a single Virasoro algebra yields the boundary Carrollian conformal algebra, and a classical free-field realization in terms of string modes has been constructed. A similar contraction of the Miura realization appears feasible and could lead to a definition of a boundary Carrollian $W_N$-algebra, at least at the classical level. 

A key motivation for developing the free-field construction is the prospect of studying representations of Carrollian and Galilean $W_N$-algebras starting from representations of free fields. In the Galilean case, similarly to the relativistic case, one can start from highest-weight representations of the free-field algebra and thereby obtain representations of the Galilean $W_N$-algebra. See \cite{Grumiller:2014lna} for a discussion of representations of Galilean $W$-algebras and their unitarity, and \cite{Radobolja:2021nae} for a discussion of representations of the Galilean $W_3$-algebra in connection with a specific free-field realization. The Carrollian case is more subtle: its representation theory differs from the Galilean one, but some of the representations discussed in the literature -- such as the induced representations of the Carrollian $W_3$-algebra in~\cite{Campoleoni:2015qrh,Campoleoni:2016vsh} -- should nevertheless be realizable in terms of free fields. This suggests that analogous constructions could be extended to higher Carrollian $W_N$-algebras. It would also be interesting to analyze correlation functions (see, e.g.\ \cite{Bagchi:2009ca,Bagchi:2009pe}) and the role of screening charges in the Galilean or Carrollian setting.

Finally, an important open question is whether dynamical field theories with Carrollian $W_N$-symmetry can be constructed. In the relativistic case, conformal Toda theories provide canonical examples of $W_N$-symmetric field theories. A Carrollian analog of conformal Toda theory would offer fertile ground to investigate the dynamics associated with these extended symmetries (see for example~\cite{Gonzalez:2014tba} for developments in this direction). We hope that our systematic analysis of Carrollian $W_N$-algebras will contribute to a better understanding of quantum theories with extended Carrollian symmetry.
\section*{Acknowledgements}
\label{sec:acknowledgement}
We thank Paulina Schlachter for initial collaboration on this project. We also thank Andrea Campoleoni and José Figueroa-O’Farrill for helpful discussions. 

\appendix
\section*{Appendix}
\section{Galilean contraction of $W_{N}$-algebras}
\label{app:Galilean}

A well-known result is that a Galilean contraction of two copies of the Virasoro algebra,
\begin{align}
    \hat{\mL}_m &= L_m + \bar{L}_m \,,& \hat{\mM}_m &=\epsilon^2 \big(L_m - \bar{L}_m\big)\,,
    \label{GalileanContraction}
\end{align}
leads to the same commutation relations~\eqref{CCA} as those obtained from the Carrollian contraction with central charges $\hat{c}_L=c+\bar{c}$ and $\hat{c}_M= \epsilon^2(c-\bar{c})$. Similarly to Section \ref{sec:Miura}, we can also investigate the contraction of the Miura transformation from the Galilean perspective. The resulting Galilean $W_N$-algebras coincide with the Carrollian ones. Concretely, one can consider the combinations
\begin{subequations}\label{GalileanCompatibleFields}
\begin{align}
    \hat{\mU}^{+}_k &= \frac{\epsilon^k}{2}\big(u_k + i^{-k}\bar{u}_k\big), & \hat{\mU}^-_k &= \frac{\epsilon^{k-2}}{2}\big(u_k - i^{-k}\bar{u}_k\big)\,,\\
    \hat{\mJ}^+_j &= \frac{\epsilon}{2}\big( J_j - i \bar{J}_j\big), & \hat{\mJ}^-_j &= \frac{\epsilon^{-1}}{2}\big( J_j + i \bar{J}_j\big)\,,\\
    \hat{\alpha}^+ &= \frac{\epsilon}{2}\big( \alpha - i \bar{\alpha}\big), & \hat{\alpha}^- &= \frac{\epsilon^{-1}}{2}\big( \alpha + i \bar{\alpha}\big)\,.
\end{align}
\end{subequations}
Inverting these relations leads to 
\begin{subequations}
\begin{align}
    u_k & = \epsilon^{-k}\hat{\mU}^+_k + \epsilon^{2-k} \hat{\mU}^-_k, & \bar{u}_k &= i^k \big(\epsilon^{-k} \hat{\mU}^+_k - \epsilon^{2-k} \hat{\mU}^-_k\big),\\
    J_j &= \epsilon^{-1} \hat{\mJ}^+_j + \epsilon \hat{\mJ}^-_j, & \bar{J}_j &= i\big(\epsilon^{-1} \hat{\mJ}^+_j - \epsilon \hat{\mJ}^-_j\big),\\
    \alpha &= \epsilon^{-1}\hat{\alpha}^+ + \epsilon \hat{\alpha}^-, & \bar{\alpha} &= i\big(\epsilon \hat{\alpha}^+ - \epsilon \hat{\alpha}^-\big)\,.
\end{align}
\end{subequations}
Then a single factor in the Miura transformation can be rewritten as
\begin{subequations}
\begin{align}
    \alpha \partial - J_i &= \epsilon^{-1} \big( (\alpha^+ \partial - \mJ^+_i) + \epsilon^2(\alpha^- \partial -\mJ^-_i)\big)\,,\\
    \bar{\alpha} \partial - \bar{J}_i &= i\epsilon^{-1} \big( (\alpha^+ \partial - \mJ^+_i) - \epsilon^2(\alpha^- \partial -\mJ^-_i)\big)\,.
\end{align}
\end{subequations}
Similarly, for the right-hand sides of the Miura transform, we find
\begin{subequations}
\begin{align}
    (\alpha \partial)^N - \sum_k u_k (\alpha \partial)^{N-k} &= \epsilon^{-N}\Big\{  \big( (\alpha^+ + \epsilon^2 \alpha^-)\partial \big)^N \nonumber\\
    & \qquad\qquad- \sum_k \big(\hat{\mU}^+_k + \epsilon^2 \hat{\mU}^-_k\big)\big( (\alpha^+ +\epsilon^2 \alpha^-)\partial \big)^{N-k} \Big\}\,,\label{GalileanHolomMiura}
    \\
     (\bar{\alpha} \partial)^N - \sum_k \bar{u}_k (\bar{\alpha} \partial)^{N-k} &= i^N\epsilon^{-N}\Big\{  \big( (\alpha^+ - \epsilon^2 \alpha^-)\partial \big)^N \nonumber\\
     &\qquad\qquad- \sum_k \big(\hat{\mU}^+_k - \epsilon^2 \hat{\mU}^-_k\big)\big( (\alpha^+ -\epsilon^2 \alpha^-)\partial\big)^{N-k} \Big\}\,.\label{GalileanAntiholomMiura}
\end{align}
\end{subequations}
From these expressions, one observes that the Galilean contraction leads to the same relations between $\hat{\mU}^\pm_k$ and $\hat{\mJ}^\pm_j$ in the limit $\epsilon \to 0$ as in the Carrollian case (see~\eqref{CarrollMT1} and~\eqref{CarrollMT2a}). One can also check the expressions for the central charges:
\begin{align}
    \hat{c}_L &= c+ \bar{c}\\
    &= (N-1)\big(1- N(N+1)\alpha^2\big) + (N-1) \big(1-N(N+1)\bar{\alpha}^2\big)\\
    &= (N-1)\big( 2 - 4N(N+1)\alpha^+ \alpha^-\big)
\end{align}
and
\begin{align}
    \hat{c}_M &= \epsilon^2 (c-\bar{c})\\
    &= -\epsilon^2 (N-1)N(N+1) (\alpha^2 - \bar{\alpha}^2)\\
    &\xrightarrow{\epsilon\to 0} -2(N-1)N(N+1)(\alpha^+)^2\,.
\end{align}
They coincide with the expressions~\eqref{cL} and~\eqref{cM} obtained from the Carrollian contraction as expected.

The corresponding quantum algebra arises from normal-ordering of the currents. In the Galilean contraction~\eqref{GalileanContraction}, the mode numbers are not reversed, and we can obtain the Galilean fields by suitable combinations of $u_i(\tau,\theta)$ and $\bar{u}_i(\tau,-\theta)$, similar to the combinations~\eqref{FieldContractionCarroll} in the Carrollian case. This means that we can view $\bar{u}_i$ also as holomorphic fields, and the time direction (and thus the normal-ordering prescription) has not been changed. Both, \eqref{GalileanHolomMiura} and~\eqref{GalileanAntiholomMiura}, can be considered as normal-ordered, and therefore one arrives in the end at the same quantum algebra as in the flipped Carrollian construction.

\section{Anti-normal ordering}
\label{B}

In this section, we explicitly demonstrate the origin of the negative anti-normal-ordering shift in \eqref{cbar} for $N=2$. The part of the central charge that comes from normal ordering is independent of $\alpha$, so we set $\alpha=\bar{\alpha}=0$ for simplicity. We start from the expressions
\begin{align}
    L_m &= \sum_n :\!J_{m-n} J_n\!: \,,&
    \bar L_m &= \sum_n \circColon\!\bar J_{m-n} \bar J_n \!\circColon \,,
\end{align}
where the $J_n$ (and similarly $\bar{J}_n$) satisfy the commutation relations~\eqref{freefieldCR}. We denote the mode anti-normal ordering by $\circColon \ \circColon$. We now evaluate the commutator $[\bar L_m, \bar L_{-m}]$ with $m>0$ to determine the central charge:
\begin{align}
    [\bar L_m, \bar L_{-m}] &= \left[ \sum_n \circColon\!\bar J_{m-n}\bar J_n\!\circColon\,,\sum_k \circColon\!\bar J_{m-k}\bar J_k\!\circColon  \right]\\ &= \sum _n \left((m-n)\bar J_{-n} \bar J_n + n \bar J_{m-n} \bar J_{-m+n}\right)\,.
    \label{commutator-antinormal-ordering}
\end{align}
Now, we bring the quadratic expressions into anti-normal ordered form. First, we separate the sum over the index $n$ in the regions $n\leq 0$, $1\leq n \leq m$ and $n>m$. Then, we identify the relation between the terms in \eqref{commutator-antinormal-ordering} and their anti-normal ordered counterparts. In the region $n\leq 0$, the terms are already anti-normal ordered, and we can directly replace $\bar J_{-n} \bar J_n$ by $\circColon\!\bar J_{-n} \bar J_n \!\circColon$. In the region $n>m$, both terms need to be reordered, but the necessary shifts cancel. We only get a contribution from the region $1\leq n \leq m$, where only one of the terms needs to be reordered. One arrives at
\begin{align}
    [\bar L_m, \bar L_{-m}] &=  \sum_{n\leq 0} \left((m-n)\circColon\!\bar J_{-n} \bar J_n \!\circColon + n\circColon\! \bar J_{m-n} \bar J_{-m+n}\!\circColon\right) \nonumber\\ &\quad +  \sum_{n=1}^m\left((m-n)\circColon\!\bar J_{-n} \bar J_n \!\circColon + n\circColon\! \bar J_{m-n} \bar J_{-m+n}\!\circColon\right)- \frac{1}{2}\sum_{n=1}^m n(m-n) \nonumber\\ 
    &\quad+ \sum_{n>m}\left((m-n)\circColon\!\bar J_{-n} \bar J_n \!\circColon + n\circColon\! \bar J_{m-n} \bar J_{-m+n}\!\circColon\right)\,. 
\end{align}
The additional term arising from reordering evaluates to
\begin{equation}
    - \frac{1}{2}\sum_{n=1}^m n(m-n) = -\frac{1}{12}m(m^2-1)\,.
\end{equation}
The modes $\bar{L}_m$ satisfy the Virasoro algebra~\eqref{Virasoro-algebra}, and the above computation shows that the central charge is indeed negative, $\bar{c}=-1$.

\section{A free-field construction of the quantum Galilean $W_3$-algebra}
\label{sec:W3}

As an illustration of the general free-field constructions that we developed in Sections~\ref{sec:Miura} and~\ref{sec:quantum}, we work out the example of the quantum Galilean $W_3$-algebra.\footnote{In the work~\cite{Afshar:2013vka}, the classical Carrollian $W_3$-algebra has been identified as asymptotic symmetry algebra of three-dimensional gravity extended by a spin-3 gauge field. The authors also show how a quantum version of it can be obtained as a contraction of two copies of $W_3$-algebras, where they tacitly invert the normal-ordering prescription for one of the sectors corresponding to a flipped Carrollian contraction.} 
The operator product expansions (OPEs) in the following discussion have been computed for the case $N=3$ with the help of the Mathematica package \texttt{OPEdefs} \cite{Thielemans:1991uw}. For the spin-2 fields, the OPEs read\footnote{For readability we omit the tilde on the fields that we used in the flipped Carrollian construction in Subsection~\ref{sec:flipped}.}
\begin{align}
    \mT (z) \,\mT(w) &\sim  \frac{c_L}{2}\frac{1}{(z-w)^4} + \frac{2}{(z-w)^2}\mT(w) + \frac{1}{z-w} \partial \mT(w)\,,\\
    \mT(z)\,\mM(w) &\sim \frac{c_M}{2} \frac{1}{(z-w)^4} + \frac{2}{(z-w)^2}\mM(w) + \frac{1}{z-w} \partial \mM(w)\,,
\end{align}
which encode the Carrollian or Galilean conformal algebra~\eqref{CCA}.

For the spin-3 fields we build new linear combinations
\begin{subequations}\label{tildeU3pm}
\begin{align}
    \underline{\mU}^+_3 &= \mU^+_3 -\frac{\alpha^+}{2} \partial \mU^+_2 \,,\\
    \underline{\mU}^-_3 &= \mU^-_3 -\frac{\alpha^-}{2}\partial \mU^+_2 -\frac{\alpha^+}{2}\partial \mU^-_2 \,.
\end{align}
\end{subequations}
These combinations are then primary with respect to $\mT$ and satisfy the OPEs
\begin{align}
    \mT(z) \,\underline{\mU}^-_3(w) & \sim \frac{3}{(z-w)^2} \underline{\mU}^-_3 (w) + \frac{1}{z-w} \partial \underline{\mU}^-_3 (w)\,,\\
    \mT(z)\, \underline{\mU}^+_3(w) &\sim \frac{3}{(z-w)^2} \underline{\mU}^+_3 (w) + \frac{1}{z-w} \partial \underline{\mU}^+_3 (w)\,,\\
    \mM(z)\, \underline{\mU}^-_3(w) & \sim \frac{3}{(z-w)^2} \underline{\mU}^+_3 (w) + \frac{1}{z-w} \partial \underline{\mU}^+_3 (w)\,.
\end{align}
For stating the OPEs of the spin-3 fields, it is convenient to define composite spin-4 fields that are quasi-primary with respect to $\mT$,
\begin{align}
    \Theta(z) &= \,:\!\mM \mM\!:(w) \,,\\
    \Lambda (z) &= \, :\! \mT \mM \!:(w) -\frac{3}{10} \partial^2 \mM(w)  \,.
\end{align}
The OPEs of the basic primary spin-3 fields $\underline{\mU}^\pm_3$ given in~\eqref{tildeU3pm} for $N=3$ are
\begin{align}
    \underline{\mU}_3^- (z) \,\underline{\mU}^-_3 (w) &\sim -\frac{1}{(z-w)^6}\frac{(7-80 \alpha^+ \alpha^-)(\alpha^+)^2}{2}\nonumber\\
    &\quad +\frac{1}{(z-w)^4}\bigg(-\frac{5(\alpha^+)^2}{4} \mT(w)-\frac{(2-15 \alpha^+ \alpha^-)}{6} \mM(w)\bigg)\nonumber\\
    &\quad + \frac{1}{(z-w)^3} \bigg( -\frac{5(\alpha^+)^2}{8} \partial\mT(w)-\frac{(2-15 \alpha^+ \alpha^-)}{12} \partial\mM(w) \bigg)\nonumber\\
    &\quad + \frac{1}{(z-w)^2} \bigg(  \frac{1}{6}\Lambda(w) -\frac{3(\alpha^+)^2}{16} \partial^2\mT(w)-\frac{(2-15 \alpha^+ \alpha^-)}{40} \partial^2\mM(w)\bigg)\nonumber\\
    &\quad + \frac{1}{z-w} \bigg(  \frac{1}{12}\partial \Lambda(w) -\frac{(\alpha^+)^2}{24} \partial^3\mT(w)-\frac{(2-15 \alpha^+ \alpha^-)}{180} \partial^3\mM(w)\bigg)\,,
\end{align}
and
\begin{align}
    \underline{\mU}_3^- (z) \,\underline{\mU}^+_3 (w) &\sim -\frac{1}{(z-w)^6}10 (\alpha^+)^4 
    -\frac{1}{(z-w)^4}\frac{5(\alpha^+)^2}{4} \mM(w)\nonumber\\
    &\quad 
    - \frac{1}{(z-w)^3} \frac{5(\alpha^+)^2}{8} \partial\mM(w) \nonumber\\
    &\quad + \frac{1}{(z-w)^2} \bigg(  \frac{1}{12}\Theta(w) -\frac{3(\alpha^+)^2}{16} \partial^2\mM(w)\bigg)\nonumber\\
    &\quad + \frac{1}{z-w} \bigg(  \frac{1}{24}\partial \Theta(w) -\frac{(\alpha^+)^2}{24} \partial^3\mM(w)\bigg)\,.
\end{align}
To compare with the Galilean $W_3$-algebra given in~\cite{Afshar:2013vka}, we need to consider specific linear combinations of $\underline{\mU}^\pm_3$. The reason is that the authors of~\cite{Afshar:2013vka} start the contraction from the primary spin-3 field $W$ in a standard normalization that differs by a factor from the quasi-primary projection of $u_3$,\footnote{The standard normalization of $W$ is such that the leading term in the OPE is $W(z)W(w)\sim \frac{5c}{6}\frac{1}{(z-w)^6}+\dots$. The corresponding term in the OPE of the quasi-primary projection $u_3-\frac{\alpha}{2}\partial u_2$ of $u_3$ with itself has the coefficient $\frac{(4-15\alpha^2)c}{18}$ in front of the highest singularity \cite[(4.48)]{Campoleoni:2024ced}.}
\begin{equation}\label{W-vs-u3}
    W = i \bigg(\alpha^2-\frac{4}{15}\bigg)^{-\frac{1}{2}} \bigg( u_3 - \frac{\alpha}{2}\partial u_2\bigg)\,.
\end{equation}
Due to the reversed normal ordering, the corresponding expression for the field $\bar{W}$ differs slightly in the prefactor of the spin-3 field,
\begin{equation}\label{barW-vs-baru3}
     \bar{W} = i \bigg(\bar{\alpha}^2+\frac{4}{15}\bigg)^{-\frac{1}{2}} \bigg( \bar{u}_3 - \frac{\bar{\alpha}}{2}\partial \bar{u}_2\bigg)\,.
\end{equation}
To arrive at the construction of~\cite{Afshar:2013vka}, we build the combinations
\begin{align}
    \mW^- &= W-\bar{W}\,, & \mW^+ &= \epsilon^2 (W+\bar{W})\,.
\end{align}
Expressing $W$ in terms of $u_i$ (see~\eqref{W-vs-u3}) and $\bar{W}$ in terms of $\bar{u}_i$ (see~\eqref{barW-vs-baru3}), and then expressing everything in terms of $\alpha^\pm$ and $\mU^\pm_k$ via~\eqref{Carrolltransfinverted}, we find
\begin{align}
    \mW^+ &= \epsilon^2 \bigg( i \bigg(\alpha^2-\frac{4}{15}\bigg)^{-\frac{1}{2}} \bigg( u_3 - \frac{\alpha}{2}\partial u_2\bigg) + i \bigg(\bar{\alpha}^2+\frac{4}{15}\bigg)^{-\frac{1}{2}} \bigg( \bar{u}_3 - \frac{\bar{\alpha}}{2}\partial \bar{u}_2\bigg) \bigg)\\
    &\xrightarrow{\epsilon \to 0}  
     \frac{2i}{\alpha^+} \bigg( \mU_3^+ - \frac{\alpha^+}{2}\partial \mU_2^+\bigg) = \frac{2i}{\alpha^+} \underline{\mU}^+_3\,,\\
    \mW^- &=  i \bigg(\alpha^2-\frac{4}{15}\bigg)^{-\frac{1}{2}} \bigg( u_3 - \frac{\alpha}{2}\partial u_2\bigg) - i \bigg(\bar{\alpha}^2+\frac{4}{15}\bigg)^{-\frac{1}{2}} \bigg( \bar{u}_3 - \frac{\bar{\alpha}}{2}\partial \bar{u}_2\bigg) \\
    &\xrightarrow{\epsilon\to 0} \frac{i}{\alpha^+} \bigg( 2\Big(\mU_3^--\frac{\alpha^+}{2}\partial \mU^-_2 -\frac{\alpha^-}{2}\partial \mU^+_2\Big)-\Big(2\frac{\alpha^-}{\alpha^+}-\frac{4}{15(\alpha^+)^2}\Big)\Big(\mU^+_3 - \frac{\alpha^+}{2}\partial \mU^+_2\Big)\bigg)\\
    &\quad\  = \frac{2i}{\alpha^+} \bigg( \underline{\mU}^-_3 + \frac{2-15 \alpha^+ \alpha^-}{15(\alpha^+)^2} \underline{\mU}^+_3\bigg)\,.
\end{align}
Using these combinations of $\underline{\mU}^\pm_3$ we arrive at the OPEs
\begin{align}
     \mW^- (z) \,\mW^- (w) &\sim \frac{1}{(z-w)^6}\frac{5\,c_L}{6}+\frac{1}{(z-w)^4} 5\,\mT(w)+ \frac{1}{(z-w)^3} \frac{5}{2} \partial\mT(w)\nonumber\\
    &\quad + \frac{1}{(z-w)^2} \bigg(  -\frac{2}{3 (\alpha^+)^2}\Lambda(w) + \frac{-4+30\alpha^+\alpha^-}{45(\alpha^+)^4} \Theta(w) -\frac{3}{4} \partial^2\mT(w)\bigg)\nonumber\\
    &\quad + \frac{1}{z-w} \bigg(  -\frac{1}{3 (\alpha^+)^2}\partial \Lambda(w) + \frac{-2+15\alpha^+\alpha^-}{45(\alpha^+)^4} \partial \Theta(w) -\frac{1}{6} \partial^3\mT(w)\bigg)\,,
\end{align}
and 
\begin{align}
    \mW^- (z) \,\mW^+ (w) &\sim \frac{1}{(z-w)^6}\frac{5\,c_M}{6} +\frac{1}{(z-w)^4}5\,\mM(w) + \frac{1}{(z-w)^3} \frac{5}{2} \partial\mM(w)\nonumber\\
    &\quad + \frac{1}{(z-w)^2} \bigg(  -\frac{1}{3 (\alpha^+)^2} \Theta(w) +\frac{3}{4} \partial^2\mM(w)\bigg)\nonumber\\
    &\quad + \frac{1}{z-w} \bigg(  -\frac{1}{6 (\alpha^+)^2} \partial\Theta(w) +\frac{1}{6} \partial^3\mM(w)\bigg)\,.
\end{align}
The commutator algebra of the modes of the fields $\mT$, $\mM$, $\mW^\pm$ then coincides with the algebra given in~\cite{Afshar:2013vka}.

\bibliographystyle{utphys} 
\bibliography{bibl} 

\end{document}